\documentclass[12pt]{article}
\usepackage{amssymb,amsfonts,amsthm,amsmath,color}
\usepackage{graphicx,psfrag,epsf}
\usepackage{comment}
\usepackage{enumitem}
\usepackage{natbib}
\usepackage[colorlinks=true, linkcolor=blue, citecolor=blue]{hyperref}
\usepackage{multirow}
\usepackage[toc,page]{appendix}
\usepackage{mathrsfs}
\usepackage{bbm}
\usepackage{bm}
\usepackage{algorithm}
\usepackage{algpseudocode}
\usepackage{url} 
\usepackage{subcaption} 

\newcommand{\var}{\mbox{var}}

\newcommand{\dif}{{\rm d}}
\newcommand{\vz}{\boldsymbol{z}}

\newcommand{\mH}{{\mathsf H}}

\newcommand{\mW}{{\mathsf W}}

\newcommand{\mA}{{\mathsf A}}

\DeclareMathOperator*{\argmin}{argmin}

\DeclareMathOperator*{\E}{E}

\DeclareMathOperator*{\mse}{MSE}

\begin{document}

\def\spacingset#1{\renewcommand{\baselinestretch}%
{#1}\small\normalsize} \spacingset{1}

\title{Covariate Balancing Based on Kernel Density Estimates for Controlled Experiments}
\author{Yiou Li\(^{1}\), Lulu Kang\(^{2}\)\footnote{Address for correspondence: Lulu Kang, Associate Professor, Department of Applied Mathematics, Illinois Institute of Technology, Chicago, IL 60616 (E-mail: lkang2@iit.edu).}, and Xiao Huang\(^{2}\)\\
\(^{1}\)Department of Mathematical Sciences, DePaul University\\
\(^{2}\)Department of Applied Mathematics, Illinois Institute of
Technology}
\date{}
\maketitle

\begin{abstract}
Controlled experiments are widely used in many applications to investigate the causal relationship between input factors and experimental outcomes.
A completely randomized design is usually used to randomly assign treatment levels to experimental units.
When covariates of the experimental units are available, the experimental design should achieve covariate balancing among the treatment groups, such that the statistical inference of the treatment effects is not confounded with any possible effects of covariates.
However, covariate imbalance often exists, because the experiment is carried out based on a single realization of the complete randomization.
It is more likely to occur and worsen when the size of the experimental units is small or moderate. 
In this paper, we introduce a new covariate balancing criterion, which measures the differences between kernel density estimates of the covariates of treatment groups.
To achieve covariate balance before the treatments are randomly assigned, we partition the experimental units by minimizing the criterion, then randomly assign the treatment levels to the partitioned groups.
Through numerical examples, we show that the proposed partition approach can improve the accuracy of the difference-in-mean estimator and outperforms the complete randomization and rerandomization approaches.
\end{abstract}

\noindent%

\spacingset{1.5}

\section{Introduction}
The controlled experiment is a useful tool for investigating the causal relationship between experimental factors and responses.
It has broad applications in many fields, such as science, medicine, social science, business, etc.
The construction of the experimental design for a controlled experiment has two steps.
First, a factorial experimental design, that is the treatment settings of the experimental factors, is specified.  
By \emph{treatment settings} or \emph{treatment levels} we mean the combinations of the different factorial settings of all factors. 
There is a rich body of literature of classic and new methods on a factorial design \citep{wu2011experiments}, and it is beyond the scope of this paper.
The second step is to assign treatment settings to the available experimental units.
A factorial design can involve only one factor with $L$ treatment levels, or multiple factors with $L$ combinations of treatment settings, which is decided by the design. 
In either case, there are $L-1$ treatment effects, such as main effects or interactions between multiple factors, and they are commonly estimated using the difference-in-mean estimator.
Although practitioners usually use a completely randomized design for the second step, it has some limitations, as we are going to discuss next.
In this paper, we focus on the second step, how to assign experimental units to treatment levels to improve the accuracy of the difference-in-mean estimator.

In many applications, the experimental units are varied with different \emph{covariates} information.
The response measurement of an experimental unit can be influenced by both the treatment setting and the covariates information of the experimental unit.
For instance, in a clinical trial of a new hypoglycemic agent, the experimental factor is the treatment that a patient receives and it has two levels, the new agent (with a fixed dosage) and placebo. 
The experimental units are the patients who participate in the clinical trial.
The patients are partitioned into two groups. 
The group that receives the placebo is typically called the \emph{control group}, and the group that receives the new agent is called the \emph{treatment group}. 
The treatment effect of the new agent is then estimated by the difference of the average blood glucose level of the treatment group and that of the control group. 
This estimator is called \emph{mean-difference} or \emph{difference-in-mean estimator} \citep{rubin2005causal}.
The covariates of experimental units include the patients' age, gender, weight, and other physical and medical information.
Naturally, the blood glucose level of a patient (the response) is related to the treatment, as well as the physical and medical background (covariates information) of the patient.
If the distributions of covariates of the treatment and control groups are significantly different, the estimated treatment effect via the difference-in-mean estimator can be confounded with some covariate effects.
If the number of experimental units is large relative to $L$, the confounding problem can be elevated by a completely randomized design.
Asymptotically, the completely randomized design results in the same distribution of covariates for all partitioned groups.
In other words, it achieves \emph{covariates balance}.
However, in practice, the experimenter only conducts the experiment once or a few times, each time using a single realization of the complete randomization, and usually for a finite number of experimental units.
As pointed out by many existing works in the causal inference literature, relying on complete randomization can be dangerous \citep{morgan2012rerandomization, bertsimas2015power}.
Especially, when the number of experimental units is small or moderate, covariate imbalance among the partitioned groups under complete randomization could be surprisingly significant, leading to inaccurate estimates and incorrect statistical inference of treatment effects \citep{bertsimas2015power}.

The problem of inaccurate estimates induced by the covariate imbalance could be addressed by two types of methods.
One type of method is for observational data. 
The premise is that a completely randomized design is used to assign treatment levels to experimental units before data are collected.
Then some adjusting methods are applied at the data analysis stage. 
These adjusting methods include post-stratification \citep{mchugh1983post, xie2016improving}, propensity score matching \citep{rosenbaum1983the, pearl2000models} , Doubly-robust estimator \citep{funk2011doubly} , coarsened exact matching (CEM) \citep{blackwell2009cem}, etc. 
The common theme of these methods is to apply sophisticated weighting schemes to improve the original difference-in-mean estimator and reduce the mean squared error (consisting of the bias and variance) of the estimator the imbalance of covariates. 
The alternative estimators also include the least-square estimator \citep{wu2011experiments} of the treatment effects which is based on parametric model assumption. 
The literature on the observational study in Causal Inference is vast and we refer readers to \cite{imbens_rubin_2015} and \cite{rosenbaum2017observation} for a more comprehensive review.

Another kind of method aims at achieving covariate balance before the random assignment of the treatment levels \citep{kallus2018optimal} at the design stage and before data collection.
First, the experimental units are partitioned into $L$ groups according to a certain covariate balancing criterion.
Then, the $L$ treatment levels are randomly assigned to the $L$ groups.
Such methods include randomized block designs \citep{bernstein1927extension}, rerandomization \citep{morgan2012rerandomization, morgan2015rerandomization}, and the optimal partition proposed by \cite{bertsimas2015power} and \cite{kallus2018optimal}, etc.
With a randomized block design, the experimental units are divided into subgroups called blocks such that the experimental units have similar covariates information in each block.
Then, the treatment levels are randomly assigned to the experimental units within each block.
Similar to the post-stratification method, blocking can not be directly applied when the covariates are continuous or mixed.
Users need to choose a discrete block factor based on continuous or mixed covariates.
\cite{morgan2012rerandomization} and \cite{morgan2015rerandomization} proposed a rerandomization method in which the experimenter keeps randomizing the experimental units into $L$ groups to achieve a sufficiently small Mahalanobis distance of the covariates between the groups.
In \cite{bertsimas2015power}, the imbalance is measured as the sum of discrepancies in group means and variances, and the desired partition minimizes this imbalance criterion.
\cite{kallus2018optimal} proposed a new kernel allocation to divide the experimental units into balanced groups.

In this paper, we only consider the controlled experiments with continuous covariates. 
We propose a new criterion to measure the covariate imbalance, and the corresponding optimal partition minimizes this new criterion and achieves the best covariates balance between $L$ treatment groups. 
The problem set-up and assumptions are introduced in Section \ref{sec:framework}.
In Section \ref{sec:criterion}, we propose the new covariate balancing criterion, which measures the differences between the kernel density estimates of the covariates of the $L$ groups of experimental units.
In Section \ref{sec:algorithm}, we formulate the partition problem into a quadratic integer programming and discuss the choice of the parameters in kernel density estimates and optimization.
The proposed approach is compared with complete randomization and rerandomization through simulation and real examples in Section \ref{sec:examples}.
We conclude this paper with some discussions and potential research directions in Section \ref{sec:discussion}.

\section{Problem Set-up}\label{sec:framework}
In this work, we assume that the controlled experiment has $N$ experimental units and they are predetermined and fixed through the data collection process. 
We first illustrate the problem set-up using the case of $L=2$, based on which we derive the covariates balancing criterion in Section \ref{sec:criterion}. 
In the second part of Section \ref{sec:criterion}, we extend the design criterion from $L=2$ case to the general $L\geq 2$ case. 

We assume that the response variable follows a general model
\begin{equation}\label{eqn:model}
y_i = h(\bm z_i)+\alpha x_i+\epsilon_i, \,\,\,i=1,\ldots,N. 
\end{equation}
Here $x_i\in\{0, 1\}$ is the indicator of which of the two treatment levels the experimental unit $i$ recieves.
Conventionally, we use $x_i=0$ to present a baseline level, or \emph{control}, and $x_i=1$ to present the other level, or \emph{treatment}.
We loosely use the terms control and treatment just to make the distinction between the two different levels.
Assume the covariates of a experimental unit is $\bm z\in \Omega \subset \mathbb{R}^d$, and $\bm z_i = [z_{i1},\ldots,z_{id}]^{\top}$ is the observed covariates of the $i$th experimental unit.
The function $h$ is a square-integrable function.
When $x_i=0$, $h(\bm z_i)$ is the mean of the response $y_i$.
The parameter $\alpha$ is the treatment effect and $\epsilon_i$ is the random noise with zero mean and constant variance $\sigma^2$.
Furthermore, $\epsilon_i$ is independent of the covariates of all the experimental units, treatment assignment, and the random noise of other experimental units.

Based on \eqref{eqn:model}, the sample means of the response in two groups are calculated as
\[
\bar{y}_T =\frac{\sum\limits_{i=1}^N y_ix_i}{n_T}, \quad \bar{y}_C =\frac{\sum\limits_{i=1}^N y_i(1-x_i)}{n_C},
\]
where $n_T$ and $n_C$ are number of experimental units in treatment group and control group, respectively.
The most commonly used estimator for the treatment effect $\alpha$ is the difference-in-mean estimator
\begin{eqnarray}\label{eqn:est}
\hat{\alpha} &=& \bar{y}_T-\bar{y}_C\nonumber \\
&=& \alpha+ \int h(\bm z)\dif\hat{F}_T(\bm z)-\int h(\bm z)\dif\hat{F}_C(\bm z)+ \bar{\epsilon}_T-\bar{\epsilon}_C,
\end{eqnarray}
where $\hat{F}_T$ and $\hat{F}_C$ are the empirical distributions of the covariates of the treatment group and control group, respectively, and $\bar{\epsilon}_T= \frac{\sum\limits_{i=1}^N \epsilon_ix_i}{n_T}$ and $\bar{\epsilon}_T= \frac{\sum\limits_{i=1}^N \epsilon_i(1-x_i)}{n_C}$ are the mean errors in the corresponding groups.

Before conducting the experiment, the randomness of $\hat{\alpha}$ comes from three sources: random noise, the partition of experimental units in the two groups (if it is done through randomization), and the random assignment of two levels to the two partitioned groups.
More importantly, the three sources of randomness are independent of each other in our framework.
Accordingly, the mean of estimator $\hat{\alpha}$ is
\begin{eqnarray*}
\E(\hat{\alpha}) = {\E}_{\epsilon}\left\{\left.{\E}_{\text{Partition}}\left[{\E}_{\text{LA}}\left(\hat{\alpha}\left|\hat{F}_1,\hat{F}_2,\epsilon\right.\right)\right|\epsilon\right]\right\},
\end{eqnarray*}
where $\hat{F}_1$ and $\hat{F}_2$ are the empirical distribution of partitioned Group 1 and Group 2, respectively.
The partition can be random or deterministic, depending on the partition method used.
Define $LA$ as the Bernoulli random variable representing the Level Assignment.
Let $LA=0$ if Group 1 is assigned as the treatment group and $LA=1$ if Group 1 is assigned as the control group, and $\Pr(LA=0) = \Pr(LA=1) = \frac{1}{2}$.
Obviously, if we only consider the random level assignments given the partition and random noise,
\begin{eqnarray*}
{\E}_{\text{LA}}\left(\hat{\alpha}\left|\hat{F}_1,\hat{F}_2,\epsilon\right.\right) &=& \alpha+\Pr(LA=0)\left(\int h(\bm z)\dif\hat{F}_1(\bm z)-\int h(\bm z)\dif\hat{F}_2(\bm z)\right)\\
&&+\Pr(LA=1)\left(\int h(\bm z)\dif\hat{F}_2(\bm z)-\int h(\bm z)\dif\hat{F}_1(\bm z)\right)+ \bar{\epsilon}_T-\bar{\epsilon}_C\\
&=& \alpha+\bar{\epsilon}_T-\bar{\epsilon}_C.
\end{eqnarray*}
Since the random noise is independent of partition, regardless of distribution of the partition,
\begin{eqnarray*}
\E(\hat{\alpha}) = {\E}_{\epsilon}\left\{{\E}_{\text{Partition}}\left(\alpha+\bar{\epsilon}_T-\bar{\epsilon}_C\right)\right\} = \alpha.
\end{eqnarray*}
Therefore, it does not matter what kind of partition method we use, random or deterministic, the difference-in-mean estimator is always unbiased \citep{kallus2018optimal}, as long as the Level Assignment is fairly and randomly assigned to the two partitioned groups.

Similarly, considering the three sources of randomness, the variance of $\hat{\alpha}$ is
\begin{eqnarray}\label{eqn:variance1}
\var(\hat{\alpha}) &=& \E\left[\left(\int h(\bm z)\dif\hat{F}_T(\bm z)-\int h(\bm z)\dif\hat{F}_C(\bm z)+ \bar{\epsilon}_T-\bar{\epsilon}_C\right)^2\right]\nonumber \\
&=& \E\left[\left(\int h(\bm z)\dif\hat{F}_T(\bm z)-\int h(\bm z)\dif\hat{F}_C(\bm z)\right)^2\right]+\sigma^2\left(\frac{1}{n_T}+\frac{1}{n_C}\right).
\end{eqnarray}
In \eqref{eqn:variance1}, the expectation in the last equation is with respect to the randomness in the partition of the experimental units, and the randomness of level assignment.
We can use a more detailed but cumbersome notation and derive
\begin{align*}
&\E\left[\left(\int h(\bm z)\dif\hat{F}_T(\bm z)-\int h(\bm z)\dif\hat{F}_C(\bm z)\right)^2\right]\\
=&{\E}_{\text{Partition}}\left\{{\E}_{\text{LA}}\left[\left.\left(\int h(\bm z)\dif\hat{F}_T(\bm z)-\int h(\bm z)\dif\hat{F}_C(\bm z)\right)^2\right|\hat{F}_1,\hat{F}_2\right]\right\}\\
=& {\E}_{\text{Partition}}\left\{\Pr(LA=0)\left(\int h(\bm z)\dif\hat{F}_1(\bm z)-\int h(\bm z)\dif\hat{F}_2(\bm z)\right)^2\right.\\
&\,\,\,\,\,\,\,+\left.\Pr(LA=1)\left(\int h(\bm z)\dif\hat{F}_2(\bm z)-\int h(\bm z)\dif\hat{F}_1(\bm z)\right)^2\right\}\\
=& {\E}_{\text{Partition}}\left[\left(\int h(\bm z)\dif\hat{F}_1(\bm z)-\int h(\bm z)\dif\hat{F}_2(\bm z)\right)^2\right].
\end{align*}
As a result, once the partition of the experimental units is established, the variance of difference-in-mean estimator $\var({\hat{\alpha}})$ in \eqref{eqn:variance1} is invariant to the treatment assignment of the two groups, that is,
\begin{eqnarray}\label{eqn:variance}
\var(\hat{\alpha}) &=& {\E}_{\text{Partition}}\left[\left(\int h(\bm z)\dif\hat{F}_1(\bm z)-\int h(\bm z)\dif\hat{F}_2(\bm z)\right)^2\right]+ \sigma^2\left(\frac{1}{n_T}+\frac{1}{n_C}\right).
\end{eqnarray}
Thus, the experimenter should focus on the partition of the experimental units to reduce $\var({\hat{\alpha}})$.
In the following section, we propose an alternative partition method that aims at regulating the variance of the difference-in-mean estimator $\hat{\alpha}$.

\section{Kernel Density Estimation Based Covariate Balancing Criterion}\label{sec:criterion}

In this section, we first show the derivation of the KDE-based covariate balancing criterion for the $L=2$ following the set-up in Section \ref{sec:framework}. 
Then we extend the balancing criterion to the general $L\geq 2$ case. 

\subsection{The Case of $L=2$}

Define a partition of the experimental units as $\bm g = [g_1,\ldots,g_{N}]^{\top}$, where $g_i=0$ if the $i$th experimental unit is partitioned into Group 1 and $g_i=1$ if the $i$th experimental unit is partitioned into Group 2.
Note that $g_i$ is different from the indicator variable $x_i$ in \eqref{eqn:model}.
The order of partition and treatment level assignment does not matter.
One can partition the experimental units into two groups first and then randomly assign treatment levels.
Or, one can randomly assign treatment levels to the two groups (which are still empty), and fill the groups with experimental units afterward.
Therefore, Group 1 is not necessarily the treatment or the control group, that is, $g_i=1$ does not imply $x_i=0$ or $x_i=1$.

To construct a smooth approximation of the empirical distributions of the covariates in two groups, we estimate the corresponding distributions using the kernel density estimation.
The kernel density estimation (KDE) is a popular technique to estimate the density function of a multivariate distribution, which is a generalization of histogram density estimation but with improved statistical properties \citep{simonoff2012smoothing}.
We use this approximation for two reasons: (1) the covariates are continuous in nature, and (2) to bound $\var(\hat{\alpha})$.
With a sample $\{\vz_1,\ldots,\vz_n\}$ of a $d-$dimensional random vector drawn from a distribution with density function $f$, the kernel density estimate is defined to be
\begin{equation}\label{eqn: KDE}
\hat{f}(\bm z)=\frac{1}{n}|\mH|^{-1/2}\sum_{i=1}^n K\left(\mH^{-1/2}(\bm z-\bm z_i)\right),
\end{equation}
where $K(\cdot)$ is the kernel function which is a symmetric multivariate density function, and $\mH$ is the positive definite bandwidth matrix. With the smooth approximation of the empirical distributions, by \eqref{eqn:variance}, the variance of the difference-in-mean estimator is approximately upper bounded by
\begin{eqnarray}\label{eqn:upperbound}
\begin{aligned}
\var(\hat{\alpha}) &= \E\left[\left(\int h(\bm z)\dif\hat{F}_1(\bm z)-\int h(\bm z)\dif\hat{F}_2(\bm z)\right)^2\right]+\sigma^2\left(\frac{1}{n_T}+\frac{1}{n_C}\right) \\
&\approx \E\left[\left(\int h(\bm z)\hat{f}_1(\bm z)\dif \bm z-\int h(\bm z)\hat{f}_2(\bm z)\dif \bm z\right)^2\right]+\sigma^2\left(\frac{1}{n_T}+\frac{1}{n_C}\right)\\
&\leq \E\left[\int |h(\bm z)|^2\dif \bm z\int \left|\hat{f}_1(\bm z)-\hat{f}_2(\bm z)\right|^2\dif \bm z\right]+\sigma^2\left(\frac{1}{n_T}+\frac{1}{n_C}\right)\\
&= \|h\|_2^2\E\left[\left\|\hat{f}_1-\hat{f}_2\right\|_2^2\right]+\sigma^2\left(\frac{1}{n_T}+\frac{1}{n_C}\right),
\end{aligned}
\end{eqnarray}
where the inequality follows from Cauchy-Schwarz inequality, $\|\cdot\|_2^2$ denotes the $\mathcal{L}_2$-norm of a function, $\hat{F}_1$ and $\hat{F}_2$ are empirical distributions of covariates in Group 1 and 2, and $\hat{f}_1$ and $\hat{f}_2$ are the KDE of the covariates of the two groups, respectively.
Here the expectation is only with respect to the partition $\bm g$ as explained in \eqref{eqn:variance} in Section \ref{sec:framework}.

Since $\var(\hat{\alpha})$ depends on the function $h$, we cannot directly minimize $\var(\hat{\alpha})$ with respect to partition without making any assumption on $h$ function.
To make our approach robust to any assumption on $h$, we propose using
\begin{equation}\label{eqn:criterion}
B_{\mH}(\bm g) = \left\|\hat{f}_1-\hat{f}_2\right\|_2^2
\end{equation}
as the covariate balancing criterion to partition the experimental units.
It is the part of the upper bound that is not a constant and only depends on the partition $\bm g$.
This criterion also appeared in \cite{anderson1994two}, which proposed the same criterion as a two-sample test statistic to test whether the two samples are drawn from the same distribution. 
Their work further supports the idea of using $B_{\mH}(\bm g)$ as the partition criterion from the covariate balancing perspective. 
A small $B_{\mH}(\bm g)$ value suggests that the covariates samples in the two groups tend to be drawn from the same distribution, then the covariate information in the two groups is balanced.

To achieve a partition with small $B_{\mH}(\bm g)$, one can either find the optimal solution that minimizes $B_{\mH}(\bm g)$ or rerandomize the experimental units until a sufficiently small $B_{\mH}(\bm g)$ is obtained.
The latter way is doable when the asymptotic distribution of the criterion is available.
The asymptotic distribution of $B_{\mH}(\bm g)$ can be constructed using bootstrap method \citep{anderson1994two}.
However, different from the simple normal asymptotic distribution of Mahalanobis distance derived in \cite{morgan2012rerandomization}, due to the computation cost of the bootstrap method, calculating the threshold of $B_{\mH}(\bm g)$ is computationally expensive.
As a result, we choose to construct a partition by minimizing $B_{\mH}(\bm g)$, and we call the partition $\bm g^* = \argmin_{\bm g} B_{\mH}(\bm g)$ the KDE-based partition.
Since we use the optimization approach, our partition scheme is actually deterministic.
In other words, the partition scheme we use is to let $\bm g=\bm g^*$ with probability equal to 1. 
This discussion also applies to the general $L\geq 2$ case. 
In Section \ref{sec:algorithm}, we discuss in detail about how to construct the KDE-based partition that minimizes $B_{\mH}(\bm g)$.

\subsection{General Case of $L\geq 2$}\label{subsec:general}

For the general case of $L\geq 2$, the covariate balancing criterion is generalized as follows
\begin{equation}\label{eqn:multicriterion}
B_{\mH}(\bm g) = \max_{l,s=1,\ldots,L}\left\|\hat{f}_l-\hat{f}_s\right\|_2^2.
\end{equation}
Extending $\bm g$ to the case of $L>2$, the partition is defined as $\bm g = [g_1,\ldots,g_N]^{\top}$, with $g_i \in \{0,\ldots,L-1\}, i = 1,\ldots, N$.
To show the generalized criterion in \eqref{eqn:multicriterion} is reasonable, we explain it under two scenarios: (1) the experiment involves only one factor with $L$ treatment settings; (2) the experiment involves multiple factors and the experimental design contains $L$ different combinations of the treatment settings of these factors. 
In the first scenario, the treatment effects of interest are the pairwise contrasts between any two treatment levels. 
The linear model in \eqref{eqn:model} is extended to 
\[
y_i = h(\bm z_i)+\sum_{l=1}^{L-1}\alpha_j x_{il}+\epsilon_i, \,\,\,i=1,\ldots,N.
\]
Here $x_{il}$'s are the dummy variables corresponding to the assigned treatment level for the $i$th unit, and $x_{il}=1$ if the $i$th unit is assigned to treatment level $l$, and $x_{il}=0$ otherwise, for $l=1,\ldots, L-1$. 
Using this notation, the treatment level $L$ is set to be the baseline, and the rest of the treatment levels are compared to it. 
The treatment effects $\alpha_l$ for $l=1,\ldots, L-1$ are estimated by the difference-in-mean estimator
\[
\hat{\alpha}_l=\bar{y}_{l}-\bar{y}_L,\quad \text{for } l=1,\ldots, L-1,
\]
where $\bar{y}_k$ is the sample mean of the observations with treatment level $k$ for $k=1,\ldots, L$. 
Often, an experimenter is interested in the pairwise contrasts of the treatment effects $\alpha_l-\alpha_s$, $l,s = 1,\ldots, L$. Therefore, one would aim at minimizing the largest variance of the corresponding estimator $\hat{\alpha}_l-\hat{\alpha}_s$, that is, $\max\limits_{l,s=1,\ldots,L} \var(\hat{\alpha}_l-\hat{\alpha}_s)$.
Following the same deviation in the $L=2$ case, $\hat{\alpha}_l-\hat{\alpha}_s$ is unbiased with variance
\[
\var(\hat{\alpha}_l-\hat{\alpha}_s)={\E}_{\text{Partition}}\left[\left(\int h(\bm z)\dif\hat{F}_{l}(\bm z)-\int h(\bm z)\dif\hat{F}_s(\bm z)\right)^2\right]+ \sigma^2\left(\frac{1}{n_l}+\frac{1}{n_s}\right).
\]
Using the similar argument in \eqref{eqn:upperbound}, the largest variance is approximately upper bounded by
$$\max\limits_{l,s=1,\ldots,L} \var(\hat{\alpha}_l-\hat{\alpha}_s) \lessapprox \max\limits_{l,s=1,\ldots,L}\left\{\|h\|_2^2\E\left[\left\|\hat{f}_s-\hat{f}_l\right\|_2^2\right]+\sigma^2\left(\frac{1}{n_l}+\frac{1}{n_s}\right)\right\}.$$
Assuming all the $L$ treatment groups have very similar or the same size of experimental units, one should aim at minimizing $\max\limits_{l,s=1,\ldots,L}\left\|\hat{f}_l-\hat{f}_s\right\|_2^2$. Thus, a natural covariates balancing criterion is
\[
B_{\mH}(\bm g) = \max\limits_{l,s=1,\ldots,L}\left\|\hat{f}_l-\hat{f}_s\right\|_2^2.
\]
When $L=2$, the criterion reduces to \eqref{eqn:criterion}.


In the second scenario, we explain it via a simple $2\times 2$ full factorial design with two two-level factors, denoted by $A$ and $B$. 
The full factorial design contains $L=4$ treatment settings. 
Using the generic notation in the design of experiments literature, the four treatment settings are $(+, +)$, $(+, -)$, $(-, +)$, and $(-,-)$. 
The sample mean of the observations in each treatment group is $\bar{y}_l$ and the sample size is $n_l$ for $l=1,\ldots, L$. 
There are $L-1=3$ effects to be estimated, which are main effects of $A$ and $B$ (denoted as $\alpha_A$ and $\alpha_B$) and their interaction effect (denoted by $\alpha_{AB}$). 
For simplicity, we assume $n_1 = \ldots = n_L = n$, and the total number of experimental units is $N=4n$. 
The commonly used difference-in-mean estimators for $\alpha_A, \alpha_B$, and $\alpha_{AB}$ are 
\begin{align*}
\hat{\alpha}_A&=\frac{1}{2}\left(\bar{y}_1+\bar{y}_2-\bar{y}_3-\bar{y}_4\right)\\
\hat{\alpha}_B&=\frac{1}{2}\left(\bar{y}_1+\bar{y}_3-\bar{y}_2-\bar{y}_4\right)\\
\hat{\alpha}_{AB}&=\frac{1}{2}\left(\bar{y}_1-\bar{y}_2-\bar{y}_3+\bar{y}_4\right).
\end{align*}
Following the same derivation of \eqref{eqn:variance}, we obtain the variance of $\hat{\alpha}_A$ 
\begin{align*}
\var(\hat{\alpha}_A)&={\E}\left[\frac{1}{4}\left(\int h(\bm z)\left[\dif\hat{F}_{1}(\bm z)+\dif\hat{F}_{2}(\bm z)\right]-\int h(\bm z)\left[\dif\hat{F}_3(\bm z)+\dif\hat{F}_4(\bm z)\right]\right)^2\right]+\frac{\sigma^2}{n}.
\end{align*}
Using the argument of \eqref{eqn:upperbound}, $\var(\hat{\alpha}_A)$ is approximately upper bounded 
\begin{align*}
\var(\hat{\alpha}_A)\lessapprox \frac{1}{4}\|h\|_2^2\E\left[\left\|\hat{f}_1+\hat{f}_2-\hat{f}_3-\hat{f}_4\right\|_2^2\right]+\frac{\sigma^2}{n}.
\end{align*}
In this upper bound, only $\left\|\hat{f}_1+\hat{f}_2-\hat{f}_3-\hat{f}_4\right\|_2$ depends on the partition. 
By triangle inequality, 
\[
\left\|\hat{f}_1+\hat{f}_2-\hat{f}_3-\hat{f}_4\right\|_2\leq \min\left\{\left\|\hat{f}_1-\hat{f}_3\right\|_2+\left\|\hat{f}_2-\hat{f}_4\right\|_2, \left\|\hat{f}_1-\hat{f}_4\right\|_2+\left\|\hat{f}_2-\hat{f}_3\right\|_2\right\}. 
\]
Similarly, the corresponding norms in the upper bounds of $\var(\hat{\alpha}_B)$ and $\var(\hat{\alpha}_{AB})$ are upper bounded by
\begin{align*}
&\left\|\hat{f}_1+\hat{f}_3-\hat{f}_2-\hat{f}_4\right\|_2\leq \min\left\{\left\|\hat{f}_1-\hat{f}_2\right\|_2+\left\|\hat{f}_3-\hat{f}_4\right\|_2, \left\|\hat{f}_1-\hat{f}_4\right\|_2+\left\|\hat{f}_3-\hat{f}_2\right\|_2\right\}, \\
&\left\|\hat{f}_1+\hat{f}_4-\hat{f}_2-\hat{f}_3\right\|_2\leq \min\left\{\left\|\hat{f}_1-\hat{f}_2\right\|_2+\left\|\hat{f}_3-\hat{f}_4\right\|_2, \left\|\hat{f}_1-\hat{f}_3\right\|_2+\left\|\hat{f}_2-\hat{f}_4\right\|_2\right\}. 
\end{align*}
Therefore, to regulate the worst-case variance of these three estimators, it makes sense to minimize the largest pairwise difference $\|\hat{f}_l-\hat{f}_s\|_2$, $l,s=1,\ldots, 4$. Thus, we reach the same criterion as above
\[
B_{\mH}(\bm g) = \max_{l,s=1,\ldots,4}\left\|\hat{f}_l-\hat{f}_s\right\|_2^2.
\]

For general full or fractional factorial design, if there are $L$ treatment settings, there would be $L$ sample means of the response from each treatment group. 
If all the $L-1$ effects (main effects, two-factor interactions, etc) are parameters of interests, then all the pairwise distance of $||\hat{f}_l-\hat{f}_s||_2$ should be as small as possible so that the covariate balancing is achieved at best across all treatment groups. 
If only some of the $L-1$ effects are of interest, it is still ideal to reach covariate balancing across all treatment groups because the difference-in-mean estimators are essentially linear combinations of the group sample means. 

\section{Construction of KDE-Based Partition}\label{sec:algorithm}
Constructing a KDE-based partition of the experimental units is essentially an optimization problem.
\begin{subequations}\label{eqn:opt}
\begin{alignat}{2}
&\! \min_{\bm g} & \qquad & \max\limits_{l,s=1,\ldots,L}\left\|\hat{f}_l-\hat{f}_s\right\|_2^2\\
& \text{s.t.} & & 0\leq g_i \leq L-1,\,\,\,i=1,\ldots,N,\\
& & & \sum\limits_{i=1}^N \mathbbm{1}_{\{g_i=j\}} = n_j, \,\,\,j = 0,\ldots,L-1,\\
& & & g_i \,\,\,\,\text{integer}, \,\,\, i=1,\ldots,N.
\end{alignat}
\end{subequations}
Here $\mathbbm{1}_{\{\cdot\}}$ is the indicator function that $\mathbbm{1}_{\{A\}} = 1$ if $A$ is true, and $\mathbbm{1}_{\{A\}} = 0$ otherwise, and $n_j$ is the size of the experimental units in the $(j+1)$th group. This is an integer programming problem which can be difficult and computational to solve.
In the next part we formulate \eqref{eqn:opt} into a quadratic integer programming problem for $L=2$.
It can be solved efficiently using modern optimization tools.

\subsection{Optimization}
To facilitate the formulation and computation of the optimization problem, we derive a more concrete formula of $B_{\mH}(\bm g)$ defined in \eqref{eqn:multicriterion} for general $L\geq 2$. For simplicity, we assume the number of experimental units is equal for all groups, so $N$ is divisible by $L$.
Denote the number of experimental units in the $l$th partitioned group as $n$, and $N=nL$. For any $l,s=1,\ldots,L$, we have
\begin{align*}
&\left\|\hat{f}_l-\hat{f}_s\right\|^2_2=\int \left|\hat{f}_l(\bm z)-\hat{f}_s(\bm z)\right|^2\dif\bm z\\
=&\int\left(\frac{1}{n}|\mH|^{-\frac{1}{2}}\sum_{g_i=l-1} K\left({\mH}^{-\frac{1}{2}}(\bm z-\bm z_i)\right)-\frac{1}{n}|\mH|^{-\frac{1}{2}}\sum_{g_i=s-1} K\left(\mH^{-\frac{1}{2}}(\bm z-\bm z_k)\right)\right)^2\dif\bm z\\
=&\frac{1}{n^2|\mH|}\left\{\int \left[\sum_{g_i=l-1}K\left(\mH^{-\frac{1}{2}}(\bm z-\bm z_i)\right)\right]^2\dif\bm z+\int \left[\sum_{g_i = s-1} K\left(\mH^{-\frac{1}{2}}(\bm z-\bm z_k)\right)\right]^2\dif\bm z\right.\\
-&\left.2\int \sum_{g_i=l-1}K\left({\mH}^{-\frac{1}{2}}(\bm z-\bm z_i)\right)\sum_{g_i=s-1}K\left(\mH^{-\frac{1}{2}}(\bm z-\bm z_k)\right)\dif\bm z\right\}\\
=& \frac{1}{n^2|\mH|}\left[\textrm{sum} (\mW_{l,l})+\textrm{sum} (\mW_{s,s})-2\textrm{sum} (\mW_{l,s})\right],
\end{align*}
where the matrix operator $\textrm{sum}(\mA)=\sum\limits_{i,j} \mA(i,j)$ is defined as the summation of all the entries of a matrix.
Then, $B_{\mH}(\bm g)$ can be computed as
\begin{eqnarray}
\label{eq:dis_W}
B_{\mH}(\bm g) = \max_{\substack{l,s=1,\ldots,L\\ l\neq s}} \frac{1}{n^2|\mH|}\left[\textrm{sum} (\mW_{l,l})+\textrm{sum} (\mW_{s,s})-2\textrm{sum} (\mW_{l,s})\right],
\end{eqnarray}
where the matrix $\mW$ is a symmetric matrix of size $N\times N$, with elements defined as,
\begin{align}
\label{eq:Wii}
\mW(i,i)&=\int K\left(\mH^{-1/2}(\bm z-\bm z_i)\right)^2d\bm z,\\
\label{eq:Wij}
\mW(i,j)&=\mW(j,i)=\int K\left(\mH^{-1/2}(\bm z-\bm z_i)\right)K\left(\mH^{-1/2}(\bm z-\bm z_j)\right)d\bm z.
\end{align}
It is well-known that the choice of kernel function $K$ is not essential to the KDE \citep{silverman1986density}. 
To illustrate the partition method, we choose the commonly used multivariate Gaussian kernel $K(\bm x)=(2\pi)^{-d/2}\exp(-\frac{1}{2}\bm x'\bm x)$. 
The entries of $\mW$ using the Gaussian kernel can be calculated analytically,
\begin{eqnarray*}
&& \mW(i,j)\\
&=& \int_{\mathbb{R}^d} K\left(\mH^{-\frac{1}{2}}(\vz-\vz_i)\right)K\left(\mH^{-\frac{1}{2}}(\vz-\vz_j)\right)\dif \vz\\
&=&\int_{\mathbb{R}^d} (2\pi)^{-d}\exp\left(-\frac{1}{2}\left[(\vz-\vz_i)'\mH^{-1}(\vz-\vz_i)+(\vz-\vz_j)'\mH^{-1}(\vz-\vz_j)\right]\right)\dif \vz\\
&=&(2\pi)^{-d}\int_{\mathbb{R}^p} \hspace{-1ex}\exp\left(-\left(\vz-\frac{\vz_i+\vz_j}{2}\right)'\mH^{-1}\left(\vz-\frac{\vz_i+\vz_j}{2}\right)-\frac{1}{4}(\vz_i-\vz_j)'\mH^{-1}(\vz_i-\vz_j)\right)\dif \vz\\
&=& 2^{-d}\pi^{-\frac{d}{2}}|\mH|^{\frac{1}{2}}e^{\left(-\frac{1}{4}(\vz_i-\vz_j)'\mH^{-1}(\vz_i-\vz_j)\right)}.
\end{eqnarray*}
As a special case, $\mW(i,i)= |\mH|^{\frac{1}{2}}2^{-d}\pi^{-\frac{d}{2}}$.
Note that the above calculation applies when the domain of $\bm z$, denoted by $\Omega$, is unbounded, i.e., $\Omega =\mathbb{R}^d$.
If $\Omega$ is a subset of $\mathbb{R}^d$, we can derive the integration in the range of $\Omega$, and the resulting formula would involve the CDF of normal.
But here we still integrate with the range of $\mathbb{R}^d$, and the approximation error is small since the value of the estimated density function should be small outside of $\Omega$.

Given a specific partition $\bm g$, we partition the $\mW$ matrix into $L\times L$ sub-matrices accordingly, such that each sub-matrix $\mW_{s,r}$ corresponds to the experimental units in group $r$ and $s$.
That is, the entries of sub-matrix $\mW_{r,s}$ are $\mW(i,j)$ such that $g_i=r-1$ and $g_j=s-1$ for $r,s=1,\ldots,L$.
Notice that such a definition of the block matrices depends on the partition $\bm g$.
Thus, the entries of the sub-matrices would change as the partition is varied.
But the entries of the $\mW$ for each pair of $(\bm z_i, \bm z_j)$ remain the same for all $i, j=1,\ldots,N$.
So the entries of the matrix $\mW$ only need to be computed once for computing $B_{\mH}(\bm g)$ values for different partitions.

For $L=2$, the objective function $B_{\mH}(\bm g) = \frac{1}{n^2|\mH|}\left[\textrm{sum} (\mW_{1,1})+\textrm{sum} (\mW_{2,2})-2\textrm{sum} (\mW_{1,2})\right]$. Recall that by the definition of partition vector $\bm g = [g_1,\ldots,g_N]^{\top}$, $g_i=0$ if the $i$th experimental unit is in Group 1, and $g_i=1$ if the $i$th experimental unit is in Group 2.
Then, the objective function $B_{\mH}(\bm g)$ could be rewritten as
\begin{eqnarray}
B_{\mH}(\bm g) &=& \sum_{i=1}^N\sum_{j=1}^N (2g_i-1)\mW(i,j)(2g_j-1) \nonumber\\
&=& 4\left[\sum_{i=1}^N\sum_{j=1}^N g_ig_j\mW(i,j)-\sum_{i=1}^N g_i\sum_{j=1}^N\mW(i,j)\right]+\sum_{i=1}^N\sum_{j=1}^N \mW(i,j)\nonumber\\
&=& 4(\bm g^{\top} \mW\bm g-\bm g^{\top} \bm w)+\sum_{i=1}^N\sum_{j=1}^N \mW(i,j),
\end{eqnarray}
where $\bm w = \left[\sum_{j=1}^N\mW(1,j),\ldots,\sum_{j=1}^N\mW(N,j)\right]^{\top}$. As a result, for $L=2$, the optimization problem \eqref{eqn:opt} is reformulated into
\begin{subequations}\label{eqn:optprob}
\begin{alignat}{2}
&\! \min_{\bm g} & \qquad & \bm g^{\top} \mW\bm g-\bm g^{\top} \bm w\\
& \text{s.t.} & & \sum\limits_{i=1}^N g_i = N/2, \\
& & & g_i \,\,\,\,\text{binary}, \,\,\, i=1,\ldots,N,
\end{alignat}
\end{subequations}

This is a quadratic integer programming that can be solved efficiently by Gurobi Optimizer \citep{gurobi} for small- or moderately-sized experiments.
For large-sized experiments or the more general case of $L\geq 3$, stochastic optimization tools such as genetic algorithm \citep{brad95} and simulated annealing \citep{van1987simulated} can be adopted to solve the optimization.
Regardless of the optimization method, the matrix $\mW$ is computed only once in the optimization procedure, which significantly cuts down the computation.

\subsection{Choice of bandwidth matrix}
The accuracy of the KDE is sensitive to the choice of bandwidth matrix $\mH$ \citep{wand93,sim96}.
Many methods have been developed to construct $\mH$ under various criteria \citep{sheather1991reliable, sain1994cross, wand1994multivariate, jones96, duong2005cross, zhang2006bayesian, de2011bayesian}.
Besides these methods, $\mH$ can be chosen by some rules of thumb, including Silverman's rule of thumb \citep{silverman86} and Scott's rule \citep{scott92}.
But they may lead to a suboptimal KDE \citep{duong03,wand93} due to the diagonality constraint of $\mH$.
Another rule of thumb uses a full bandwidth matrix as
\begin{eqnarray}\label{eqn:bandwidth}
\mH = n^{-2/(d+4)}\hat{\Sigma},
\end{eqnarray}
where $\hat{\Sigma}$ is the estimated covariance matrix with sample size $n$, and $d$ is the covariate dimension.
It can be considered as a generalization of Scott's rule \citep{hardle04}.
To compromise between the computational cost and the accuracy of the KDE, we propose using the rule of thumb in \eqref{eqn:bandwidth}.
It is easy to compute and leads to a more accurate KDE compared to the simple diagonal matrix.
We did a series of numerical comparisons to test the impact of $\mH$ on the KDE-based partition (not reported here due to the space limit).
The results show that \eqref{eqn:bandwidth} leads to similar partitions compared with the more computationally demanding methods such as cross-validation \citep{sain1994cross,duong2005cross} and Bayesian methods \citep{zhang2006bayesian,de2011bayesian}. 
Similar observations were found in the work by \cite{anderson1994two}. 
As pointed out in \cite{anderson1994two}, the criterion $B_{\mH}(\bm g)$ aims at measuring the discrepancy between the distributions from which the samples are drawn, but not precisely estimating those distributions. 
Although the bandwidth matrix $\mH$ plays an important role in the estimation of distribution, it is not surprising that the KDE-based partition is robust to the choice of $\mH$.

\section{Examples}\label{sec:examples}
{\bf Example 1. Simulation Example.} We compare the performance of the proposed KDE-based partition with complete randomization and rerandomization \citep{morgan2012rerandomization} through a simulation example with $d=2$ covariates. 
Three different types of mean function $h$ are considered.
\begin{itemize}
\item []\textbf{Model 1. Linear basis}: $h(\bm z) = \beta_0+\sum\limits_{i=1}^2\beta_i z_i$,
$$y = \alpha x+\beta_0+\sum\limits_{i=1}^2\beta_i z_i+\epsilon.$$

\item [] \textbf{Model 2. Quadratic basis}: $h(\bm z) = \beta_0+\sum\limits_{i=1}^2\beta_i z_i+\sum\limits_{i=1}^2\gamma_i z_i^2+\theta z_1z_2$,
$$y = \alpha x+\beta_0+\sum\limits_{i=1}^2\beta_i z_i+\sum\limits_{i=1}^2\gamma_i z_i^2+\theta z_1z_2+\epsilon.$$

\item []\textbf{Model 3 Sinusoidal model}: $h(\bm z) = \beta_0+\beta_1\sin\left(\phi+\pi\gamma_1z_1+\pi\gamma_2z_2\right)$,
$$y = \alpha x + \beta_0+\beta_1\sin\left(\phi+\pi\gamma_1z_1+\pi\gamma_2z_2\right)+\epsilon.$$
\end{itemize}
The notation $x\in \{0,1\}$ is the indicator of the treatment level assignment.
To generate data from these models, we need to specify the values of the parameters, including the treatment effect $\alpha$, and others $\beta_i$'s, $\gamma_i$'s, $\phi$ and $\theta$.
Let $\alpha = 2$ for all models.
In Model 1 and Model 2, the regression coefficients $\beta_i$'s, $\gamma_i$'s and $\theta$ are sampled from a uniform distribution $U[-2,2]$.
In Model 3, $\beta_0$, $\beta_1$, $\gamma_1$ and $\gamma_2$ are randomly generated from $U[-1,1]$, and $\phi$ is randomly generated from $U[0,2\pi]$.
The observed covariates are generated from multivariate standard normal distribution $N(\mathbf{0},\bm I_2)$.
All these values are fixed through the simulations.
Since $\var(\hat{\alpha})$ is only affected by the partition methods and is invariant to the variance of the noise $\sigma^2$, it does not matter to the comparison of different methods. 
Therefore in this and the next example, we set $\sigma=0$.

\begin{figure}[htbp]
\centering
\includegraphics[width=\linewidth]{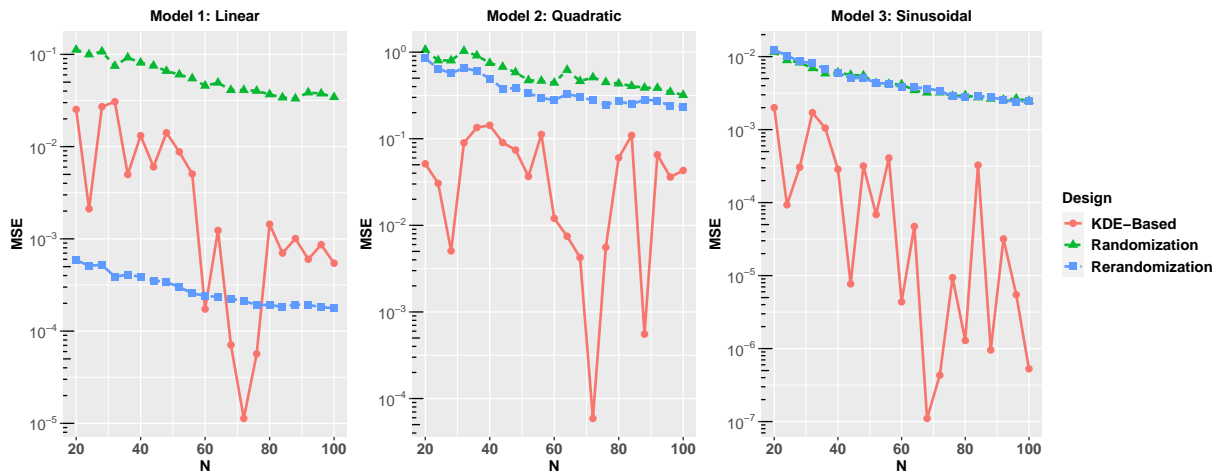}
\caption{Comparison of the estimated mean squared error of difference-in-mean estimator using three partition methods for Example 1. }
\label{fig:Ex1}
\end{figure}

To compare the proposed approach with other random methods, $m=1000$ random partitions are generated via complete randomization and rerandomization approaches.
For each partition, the treatment levels are randomly assigned to the two groups, and $m=1000$ designs are generated for the two random approaches.
Since the proposed KDE-partition is an optimization-based method, the partition is deterministic.
With all possible treatment level assignments, there are only two designs.
For each design obtained from the three methods, we generate the response data from the three models with the fixed parameters and covariates.
Then, the estimated mean squared error $\widehat{\mse}(\hat{\alpha}) = \frac{1}{m}\sum_{i=1}^{m} (\alpha-\hat{\alpha})^2$ of the difference-in-mean estimator is calculated for increasing sample size from $N = 20$ to $N=100$, and they are plotted in Figure \ref{fig:Ex1}.

When the true relationship between the response $y$ and covariate $\bm z$ is linear, rerandomization outperforms the other two methods.
\cite{morgan2012rerandomization} showed that, compared to complete randomization, the rerandomization with Mahalanobis distance criterion can reduce $\var(\hat{\alpha})$ significantly when the mean function $h(\bm z) = \sum\limits_{i=0}^d \beta_jz_j$ contains only the linear terms of the covariates.
When a more complicated relationship such as Model 2 or 3 is considered, the KDE-based partition outperforms the complete randomization and rerandomization by a large margin.
In practice, the true mean function $h$ is rarely as simple as a linear function and usually contains higher-order terms of the covariates.
Thus, from a practical perspective, we suggest that the KDE-based partition is a better choice.

We further explore the performance of the proposed KDE-based partition in matching the empirical distributions of the covariates in two groups.
We compare the difference of the empirical distributions under complete randomization, rerandomization, and KDE-based partition.
The discrepancy of the first and second raw moments of the two empirical distributions over the $m=1000$ partitions are calculated and reported in Table \ref{tab:MomentDiscEx1}.
In general, rerandomization performs the best in matching the means of the empirical distributions, which to some extent implies that it performs the best under \textbf{Model 1} (the model with main effect only).
The KDE-based partition consistently outperforms the complete randomization and is superior to the other two methods for the second moments since it matches the approximated density functions rather than just the first and second moments.

\begin{table}[htbp]
\centering
\caption{Discrepancy of Moments under Different Partition Methods}
\begin{tabular}{llccccc}
\hline
\multirow{2}{*}{$N$} & \multirow{2}{*}{Method} & & & Moments & & \\
\cline{3-7}
& & $z_1$ & $z_2$ & $z_1^2$ & $z_2^2$ & $z_1z_2$ \\
\hline
20 & Random & 0.356 & 0.298 & 0.605 & 0.292 & 0.217 \\
& Re-random & 0.026 & 0.021 & 0.612 & 0.286 & 0.235 \\
& KDE-based & 0.107 & 0.179 & 0.390 & 0.082 & 0.010 \\
40 & Random & 0.261 & 0.235 & 0.356 & 0.212 & 0.249 \\
& Re-random & 0.019 & 0.017 & 0.366 & 0.233 & 0.258 \\
& KDE-based & 0.177 & 0.011 & 0.097 & 0.071 & 0.272 \\
60 & Random & 0.204 & 0.182 & 0.278 & 0.181 & 0.189 \\
& Re-random & 0.015 & 0.014 & 0.282 & 0.188 & 0.192 \\
& KDE-based & 0.048 & 0.040 & 0.036 & 0.006 & 0.164 \\
80 & Random & 0.170 &0.177 & 0.234 & 0.201 & 0.164 \\
& Re-random & 0.013 &0.013 & 0.238 & 0.210 & 0.163 \\
& KDE-based & 0.031 & 0.122 & 0.182 & 0.144 & 0.003 \\
100 & Random & 0.167 & 0.150 & 0.238 & 0.193 & 0.161 \\
& Re-random & 0.013 & 0.011 & 0.244 & 0.198 & 0.158 \\
& KDE-based & 0.079 & 0.063 & 0.136 & 0.180 & 0.068 \\
\hline
\end{tabular}%
\label{tab:MomentDiscEx1}%
\end{table}%

{\bf Example 2. Real Data Example.} We compare the KDE-based partition with the complete randomization and rerandomization using a real data set.
The data set is from a diabetes study \citep{efron2004least}.
It contains 422 observations of $d=10$ covariates and a univariate response.
The covariates are age, sex, body mass index, average blood pressure, and six blood serum measurements of the patients, and the response is a quantitative measure of disease progression.
The data are only observational data and do not contain any experimental factors.

To use this data, we do not assume any functional form for $h(\bm z)$.
Instead, we assume the observed quantitative measure of disease progression is the sum $h(\bm z)+\epsilon$.
Let the true value of the treatment effect $\alpha=2$.
Given treatment assignment $x$, the response data including the treatment effect is, $y=\alpha x+h(\bm z)+\epsilon$.

\begin{figure}[htbp]
\centering
\includegraphics[width=10cm]{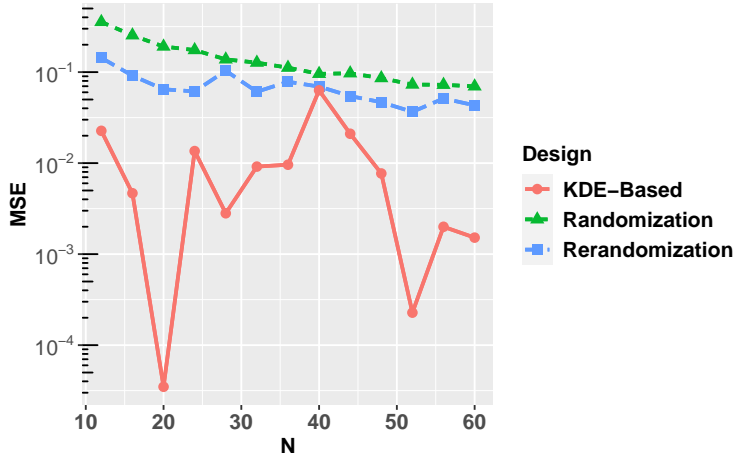}
\caption{Comparison of the estimated mean squared error of difference-in-mean estimator using three partition methods for Example 2.}
\label{fig:Ex2}
\end{figure}

For different values of $N$, ranging from $N=12$ to $N=60$, $m=1000$ partitions are generated using complete randomization and rerandomization.
As explained before, for each $N$ value, the optimal KDE-based partition is deterministic.
For each of the partitions obtained from the three methods, we randomly assign treatment settings to the two partitioned groups to obtain the design, and then compute the response $y$ accordingly.
The estimated mean squared error $\widehat{\mse}(\hat{\alpha}) = \frac{1}{m}\sum_{i=1}^{m} (\alpha-\hat{\alpha})^2$ of the difference-in-mean estimator is calculated for each $N$.
They are shown in Figure \ref{fig:Ex2}.
The KDE-based partition outperforms the other two partition methods for all sample sizes. 

\section{Discussion}\label{sec:discussion}
In this paper, we introduce a KDE-based partition method for the controlled experiments.
By adopting a smooth approximation of the covariate empirical distributions, we propose a new covariate balancing criterion. 
It measures the difference between the distributions of covariates in the partitioned groups.
We use quadratic integer programming to construct the partition that minimizes the covariate balancing criterion for the two-level experiments.
If the number of treatment settings is more than two, other stochastic optimization methods can be applied.
The design generated via the KDE-partition can regulate the variance of the difference-in-mean estimator.
Compared with the complete randomization and rerandomization methods, the simulation and real examples show that the proposed method leads to a more accurate difference-in-mean estimation of the treatment effect when the underlying model involves more complicated functions of the covariates.
The simulation example also confirms that the covariates' distributions of the groups are better matched using the proposed method.

It is worth pointing out that, when the KDE-based partition is used, the classical hypothesis testing procedure for the difference-in-mean estimator is not applicable, since the partition is a deterministic solution and the random treatment assignments only provide two different designs when $L=2$.
Fortunately, a sophisticated testing procedure using bootstrap method has been established and proven to be powerful \citep{bertsimas2015power} for the sharp null hypothesis \citep{rubin1980randomization}, $H_0: \text{all treatment effects are }0$. 
The detailed bootstrap algorithm is in Algorithm 1. 
\begin{algorithm}[ht]
\caption{Hypothesis testing procedure}
\begin{algorithmic}[1]
\Procedure {Hypothesis testing of treatment effect}{}
\State Construct KDE-based partition, randomly assign treatment levels to treatment groups, apply treatments, measure the responses $y_i$, $i=1,\ldots,N$ and compute the difference-in-mean estimate $\hat{\alpha}$.
\For {$t=1,...,T$}
\State sample $i_j^t\sim$ unif $(1,\ldots,N)$ independently for $j=1,\ldots,N$,
\State construct KDE-based partition for $\bm z_{i_1^t},\ldots,\bm z_{i_N^t}$ and
\State compute the new difference-in-mean estimate $\hat{\alpha}^t$
\EndFor
\State the $p$-value of $H_0$ is $p = \frac{1+\sum_{t=1}^T \mathbbm{1}_{\{|\hat{\alpha}^t|\geq|\hat{\alpha}|\}}}{1+T}$.
\EndProcedure
\end{algorithmic}
\end{algorithm}

One limitation of the proposed KDE-based partition is the ``curse of dimensionality''. It is well-known that KDE may perform poorly when the dimension of the covariate is large relative to the sample size.  
To overcome this problem, the experimenter can add a dimension reduction step prior to the partition of the experimental units. Based on the properties of the covariate data, one can choose the appropriate dimension reduction method from various choices, such as the principal component analysis (PCA) and the nonlinear variants of PCA. 
For instance, using PCA, the experimenter can select a small but sufficient number of the principal components and apply the KDE-based partition on the linearly transformed covariates of a much lower dimension. 
We also want to alert the readers with another limitation of the KDE-based partition. 
Similar to other optimal covariates balancing ideas, such as \cite{bertsimas2015power} and \cite{kallus2018optimal}, the KDE-based partition assumes all the influential covariates to the response are known to the experimenter and their data are included in the observed covariates data. 
Otherwise, if there are latent but important covariates, the optimal partition methods, including the proposed KDE-based partition, might lead to an estimator with large variance because it deterministically balances the experimental units based on incomplete covariate information. 
In this case, we recommend the randomization or rerandomization methods. 

The proposed KDE-based partition method can be used in other scenarios beyond controlled experiments. 
Essentially, we have proposed a density-based partition method that minimizes the differences of data between groups.
It can be incorporated into any statistical tool that needs to partition data into similar groups, such as cross-validation, divide-and-conquer, etc.
We hope to explore these directions in the future.
In this work, we do not assume any interaction terms between the covariates and the treatment effect.
However, interaction effects are likely to occur in practice.
Another interesting direction is the partition of the experimental units considering the interaction terms in the model. 

\bibliography{Ref_ABtesting}

\begin{thebibliography}{36}
\newcommand{\enquote}[1]{``#1''}
\expandafter\ifx\csname natexlab\endcsname\relax\def\natexlab#1{#1}\fi

\bibitem[{Anderson et~al.(1994)Anderson, Hall, and
  Titterington}]{anderson1994two}
Anderson, N.~H., Hall, P., and Titterington, D.~M. (1994), \enquote{Two-sample
  test statistics for measuring discrepancies between two multivariate
  probability density functions using kernel-based density estimates,}
  \textit{Journal of Multivariate Analysis}, 50, 41--54.

\bibitem[{Bernstein(1927)}]{bernstein1927extension}
Bernstein, S. (1927), \enquote{Sur l'extension du th{\'e}or{\`e}me limite du
  calcul des probabilit{\'e}s aux sommes de quantit{\'e}s d{\'e}pendantes,}
  \textit{Mathematische Annalen}, 97, 1--59.

\bibitem[{Bertsimas et~al.(2015)Bertsimas, Johnson, and
  Kallus}]{bertsimas2015power}
Bertsimas, D., Johnson, M., and Kallus, N. (2015), \enquote{The power of
  optimization over randomization in designing experiments involving small
  samples,} \textit{Operations Research}, 63, 868--876.

\bibitem[{Blackwell et~al.(2009)Blackwell, Iacus, King, and
  Porro}]{blackwell2009cem}
Blackwell, M., Iacus, S., King, G., and Porro, G. (2009), \enquote{cem:
  Coarsened exact matching in Stata,} \textit{The Stata Journal}, 9, 524--546.

\bibitem[{de~Lima and Atuncar(2011)}]{de2011bayesian}
de~Lima, M.~S. and Atuncar, G.~S. (2011), \enquote{A Bayesian method to
  estimate the optimal bandwidth for multivariate kernel estimator,}
  \textit{Journal of Nonparametric Statistics}, 23, 137--148.

\bibitem[{Duong and Hazelton(2003)}]{duong03}
Duong, T. and Hazelton, M. (2003), \enquote{Plug-in bandwidth matrices for
  bivariate kernel density estimation,} \textit{Journal of Nonparametric
  Statistics}, 15, 17--30.

\bibitem[{Duong and Hazelton(2005)}]{duong2005cross}
Duong, T. and Hazelton, M.~L. (2005), \enquote{Cross-validation Bandwidth
  Matrices for Multivariate Kernel Density Estimation,} \textit{Scandinavian
  Journal of Statistics}, 32, 485--506.

\bibitem[{Efron et~al.(2004)Efron, Hastie, Johnstone, Tibshirani,
  et~al.}]{efron2004least}
Efron, B., Hastie, T., Johnstone, I., Tibshirani, R., et~al. (2004),
  \enquote{Least angle regression,} \textit{The Annals of statistics}, 32,
  407--499.

\bibitem[{Funk et~al.(2011)Funk, Westreich, Wiesen, St{\"u}rmer, Brookhart, and
  Davidian}]{funk2011doubly}
Funk, M.~J., Westreich, D., Wiesen, C., St{\"u}rmer, T., Brookhart, M.~A., and
  Davidian, M. (2011), \enquote{Doubly robust estimation of causal effects,}
  \textit{American journal of epidemiology}, 173, 761--767.

\bibitem[{Gurobi~Optimization(2020)}]{gurobi}
Gurobi~Optimization, L. (2020), \enquote{Gurobi Optimizer Reference Manual,} .

\bibitem[{H{\"a}rdle et~al.(2012)H{\"a}rdle, M{\"u}ller, Sperlich, and
  Werwatz}]{hardle04}
H{\"a}rdle, W.~K., M{\"u}ller, M., Sperlich, S., and Werwatz, A. (2012),
  \textit{Nonparametric and semiparametric models}, New York, NY, USA: Springer
  Science \& Business Media.

\bibitem[{Imbens and Rubin(2015)}]{imbens_rubin_2015}
Imbens, G.~W. and Rubin, D.~B. (2015), \textit{Causal Inference for Statistics,
  Social, and Biomedical Sciences: An Introduction}, Cambridge University
  Press.

\bibitem[{Jones et~al.(1996)Jones, Marron, and Sheather}]{jones96}
Jones, M.~C., Marron, J.~S., and Sheather, S.~J. (1996), \enquote{Progress in
  data-based bandwidth selection for kernel density estimation,}
  \textit{Computational Statistics}, 11, 337--381.

\bibitem[{Kallus(2018)}]{kallus2018optimal}
Kallus, N. (2018), \enquote{Optimal a priori balance in the design of
  controlled experiments,} \textit{Journal of the Royal Statistical Society:
  Series B (Statistical Methodology)}, 80, 85--112.

\bibitem[{McHugh and Matts(1983)}]{mchugh1983post}
McHugh, R. and Matts, J. (1983), \enquote{Post-stratification in the randomized
  clinical trial,} \textit{Biometrics}, 217--225.

\bibitem[{Miller et~al.(1995)Miller, Goldberg, et~al.}]{brad95}
Miller, B.~L., Goldberg, D.~E., et~al. (1995), \enquote{Genetic algorithms,
  tournament selection, and the effects of noise,} \textit{Complex systems}, 9,
  193--212.

\bibitem[{Morgan and Rubin(2015)}]{morgan2015rerandomization}
Morgan, K.~L. and Rubin, D.~B. (2015), \enquote{Rerandomization to balance
  tiers of covariates,} \textit{Journal of the American Statistical
  Association}, 110, 1412--1421.

\bibitem[{Morgan et~al.(2012)Morgan, Rubin, et~al.}]{morgan2012rerandomization}
Morgan, K.~L., Rubin, D.~B., et~al. (2012), \enquote{Rerandomization to improve
  covariate balance in experiments,} \textit{The Annals of Statistics}, 40,
  1263--1282.

\bibitem[{Pearl(2000)}]{pearl2000models}
Pearl, J. (2000), \textit{Causality: Models, Reasoning, and Inference},
  Cambridge University Press.

\bibitem[{Rosenbaum(2017)}]{rosenbaum2017observation}
Rosenbaum, P. (2017), \textit{Observation and Experiment: An Introduction to
  Causal Inference}, Harvard University Press.

\bibitem[{Rosenbaum and Rubin(1983)}]{rosenbaum1983the}
Rosenbaum, P.~R. and Rubin, D.~B. (1983), \enquote{{The central role of the
  propensity score in observational studies for causal effects},}
  \textit{Biometrika}, 70, 41--55.

\bibitem[{Rubin(1980)}]{rubin1980randomization}
Rubin, D.~B. (1980), \enquote{Randomization analysis of experimental data: The
  Fisher randomization test comment,} \textit{Journal of the American
  Statistical Association}, 75, 591--593.

\bibitem[{Rubin(2005)}]{rubin2005causal}
--- (2005), \enquote{Causal inference using potential outcomes: Design,
  modeling, decisions,} \textit{Journal of the American Statistical
  Association}, 100, 322--331.

\bibitem[{Sain et~al.(1994)Sain, Baggerly, and Scott}]{sain1994cross}
Sain, S.~R., Baggerly, K.~A., and Scott, D.~W. (1994),
  \enquote{Cross-validation of multivariate densities,} \textit{Journal of the
  American Statistical Association}, 89, 807--817.

\bibitem[{Scott(2015)}]{scott92}
Scott, D.~W. (2015), \textit{Multivariate density estimation: theory, practice,
  and visualization}, Hoboken, New Jersey: John Wiley \& Sons.

\bibitem[{Sheather and Jones(1991)}]{sheather1991reliable}
Sheather, S.~J. and Jones, M.~C. (1991), \enquote{A reliable data-based
  bandwidth selection method for kernel density estimation,} \textit{Journal of
  the Royal Statistical Society. Series B (Methodological)}, 53, 683--690.

\bibitem[{Silverman(1986{\natexlab{a}})}]{silverman1986density}
Silverman, B.~W. (1986{\natexlab{a}}), \textit{Density estimation for
  statistics and data analysis}, vol.~26, CRC press.

\bibitem[{Silverman(1986{\natexlab{b}})}]{silverman86}
--- (1986{\natexlab{b}}), \textit{Density estimation for statistics and data
  analysis}, vol.~26, Boca Raton, FL: CRC press.

\bibitem[{Simonoff(2012{\natexlab{a}})}]{simonoff2012smoothing}
Simonoff, J.~S. (2012{\natexlab{a}}), \textit{Smoothing methods in statistics},
  Springer Science \& Business Media.

\bibitem[{Simonoff(2012{\natexlab{b}})}]{sim96}
--- (2012{\natexlab{b}}), \textit{Smoothing methods in statistics}, New York,
  NY: Springer Science \& Business Media.

\bibitem[{Van~Laarhoven and Aarts(1987)}]{van1987simulated}
Van~Laarhoven, P.~J. and Aarts, E.~H. (1987), \enquote{Simulated annealing,} in
  \textit{Simulated annealing: Theory and applications}, Springer, pp. 7--15.

\bibitem[{Wand and Jones(1993)}]{wand93}
Wand, M.~P. and Jones, M.~C. (1993), \enquote{Comparison of smoothing
  parameterizations in bivariate kernel density estimation,} \textit{Journal of
  the American Statistical Association}, 88, 520--528.

\bibitem[{Wand and Jones(1994)}]{wand1994multivariate}
--- (1994), \enquote{Multivariate plug-in bandwidth selection,}
  \textit{Computational Statistics}, 9, 97--116.

\bibitem[{Wu and Hamada(2011)}]{wu2011experiments}
Wu, C.~J. and Hamada, M.~S. (2011), \textit{Experiments: planning, analysis,
  and optimization}, vol. 552, Hoboken, New Jersey: John Wiley \& Sons.

\bibitem[{Xie and Aurisset(2016)}]{xie2016improving}
Xie, H. and Aurisset, J. (2016), \enquote{Improving the sensitivity of online
  controlled experiments: Case studies at netflix,} in \textit{Proceedings of
  the 22nd ACM SIGKDD International Conference on Knowledge Discovery and Data
  Mining}, ACM, pp. 645--654.

\bibitem[{Zhang et~al.(2006)Zhang, King, and Hyndman}]{zhang2006bayesian}
Zhang, X., King, M.~L., and Hyndman, R.~J. (2006), \enquote{A Bayesian approach
  to bandwidth selection for multivariate kernel density estimation,}
  \textit{Computational Statistics \& Data Analysis}, 50, 3009--3031.

\end{thebibliography}

\end{document}